# The Ambient Networks Heterogeneous Access Selection Architecture

Kostas Pentikousis[1], Ramón Agüero[2], Jens Gebert[3], José Antonio Galache[2], Oliver Blume[3], Pekka Pääkkönen[1]

[1] VTT Technical Research Centre of Finland, Kaitoväylä 1, FI-90571 Oulu, FINLAND
Tel: +358 40 536 9052. Fax: +358 20 722 2320. Email: firstname.lastname@vtt.fi
[2] Dpto. Ingeniería Comunicaciones, University of Cantabria, Avda Castros s/n 39005 Santander, SPAIN
Email: {ramon,jgalache}@tlmat.unican.es
[3] Alcatel-Lucent Deutschland AG, Research & Innovation, Lorenzstr. 10, D-70435 Stuttgart, GERMANY
Email:firstname.lastname@ alcatel-lucent.de

*Abstract*—Forthcoming wireless communications will be characterized by the ubiquity of multiaccess. Despite the inherently increased complexity, end-users should be able to take advantage of the most suitable access network. Thus, access selection in an environment with different overlapping radio technologies is of central interest and an architecture is needed that performs equally well on single- and multi-operator scenarios, considers several parameters, and respects the principle of layering. In this paper, we introduce the Ambient Networks heterogeneous access selection architecture explaining how it meets such requirements. We present the essential architectural components and explain their interactions. We illustrate how the proposed architecture works in practice and discuss recent results form our prototype-based validation.

## I. Introduction

The proliferation in radio access technologies (RATs) and industry trends indicate that, soon, even low-end devices will come with several integrated network interfaces. Most will be used in areas with overlapping coverage by several access networks and operators. Paradigms where hosts have access to various networks are not new, of course: Multihoming is used to increase resilience, dependability, and performance in high-end servers; manufacturers have been integrating different cellular RATs into "multi-band" cell phones to realize global reachability and ease migration. Nonetheless, multiaccess network selection is currently rudimentary and automation is not customary. It relies solely on presets, static choices, and user input. Multiaccess, mobility and context information does not permeate through the stack. Use of policies is limited, if existent and, basically, it is up to the user to be on the lookout for available access networks. In 3GPP reference scenarios, inter-RAT multiaccess is still an item of study.

Then, there is a rise in mobility management solutions, an increasing number of transport protocols, and an exponential growth in networked applications. In short, future networks will be more heterogeneous across all layers. This growing diversity in the entire protocol stack calls for a new design in multiaccess nodes. A design that allows dynamic use of several networks and fosters new solutions for mobility and multiaccess without hindering legacy upper layer protocols.

This paper introduces the Ambient Networks heterogeneous access selection architecture (ANHASA), which aspires to serve as a blueprint for future multiaccess and mobile systems. ANHASA is modular, incrementally deployable, and respects the principle of layering by providing generic mechanisms, not technology-specific optimizations. It employs three components of the Ambient Control Space (ACS), the overlay network control layer of the Ambient Networks project [1] and anticipates multiaccess and multioperator environments, where handovers based purely on link-layer metrics will be impossible. In §II, we introduce the architectural components along with their interactions. In §III, we show that ANHASA performs well in subsecond timescales and explain why ANHASA is larger than the sum of its components and how it enables the entire protocol stack to become cognizant of multiaccess and mobility, and play an active role in access selection. We discuss related work in §IV and conclude this paper in §V outlining future plans.

## II. ANHASA

ANHASA, the result of a layer-respecting functional decomposition, comprises three modules, which address resource abstraction, resource management, and information sharing, respectively (Fig. 1). First, the generic link layer (GLL) exposes a unified, abstract interface to all available radio accesses. Then, multiradio resource management (MRRM) uses GLL measurements and control facilities to direct access selection. Finally, trigger management (TRG) collects and distributes multiaccess and mobility information relevant to the entire protocol stack, and registers it with ANISI, the AN information service infrastructure [2].

Fig. 2 illustrates the temporal and protocol scope of GLL, MRRM, TRG, and ANISI. On the horizontal axis, we show that MRRM/GLL need to act on a sub-second timescales and TRG on up to an order of magnitude longer periods. Information of interest to entities acting on even larger timescales, such as security, mobility and context management, is registered by TRG with ANISI (see §II.D). The overall ANHASA temporal scope is in the order of one second. With respect to protocol layer scope (vertical axis), ANHASA deals with link and network layer information, letting MRRM/GLL handle events occurring in a sub-second scale and TRG/ANISI manage information relevant to events occurring at higher layers. Next, we present the ANHASA functional entities.

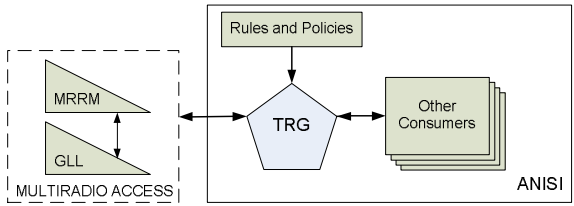

Fig. 1. ANHASA and ANISI

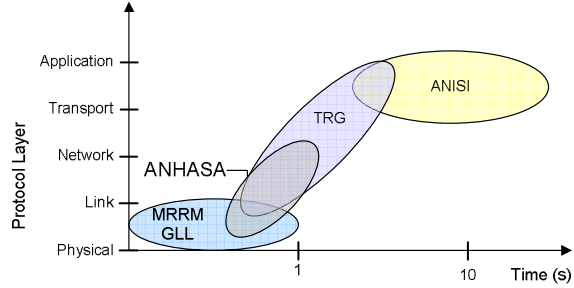

Fig. 2. ANHASA temporal and protocol layer scope

*A. Generic Link Layer*

GLL controls and delivers information from subjacent RATs and their associated networks [3][4]. Access selection expects GLL to abstract access details consistently, facilitating a fair comparison between candidates. A mapping function, which considers implementation-specific variations, supplies a "link quality" metric in the [0, 1] range, which *interprets* actual link measurements [5]. For example, when taking into account error rates, GLL considers MAC mechanisms which reduce them (residual BER). Similarly, GLL abstracts access network capacity, rate, delay, and load information. It enables resource optimization and load balancing by measuring the amount of available resources (be it codes, slots, or channels, depending on the RAT) and derives abstract metrics based on the relative resource levels. This is also used to determine if a requested QoS can be achieved.

*B. Multiradio Resource Management*

MRRM directs the advanced joint management of radio resources in heterogeneous access networks. MRRM performs access selection and load sharing between different radio accesses [5][6][7] and provisions neighborhood information. MRRM monitors for available access networks, collects link performance and resource consumption information, and correlates GLL-provided information with upper layer constraints. Based on these, MRRM takes an access selection decision, which may lead to a handover execution. Once MRRM concludes access selection, it uses GLL to establish connections to the candidate access(es). An MRRM implementation can put all functionality at the terminal only, or distribute it into subcomponents located at the terminal (MRRM-TE) and in the access network (MRRM-NET), possibly at different nodes.

*C. Trigger Management and Dissemination*

TRG allows for two-way information sharing throughout the stack, increasing the modularization of ANHASA. TRG receives events from MRRM/GLL, takes into consideration a set of rules and policies governing management and dissemination, and generates standardized notifications ("triggers"), which are delivered to subscribed functional entities and system components (called "consumers"). TRG consumers include mobility management protocols, transport protocols and applications [8]. TRG processes other dynamic information in addition to multiaccess events. It delivers triggers from upper layers to MRRM/GLL, about changes in policies, preferences, service requirements, terminal behavior and use, along with synthetic triggers based on temporal correlation of events originating from the entire protocol stack [9].

*D. ANHASA and ANISI*

Multiaccess information dissemination imposes different requirements depending on whether event details are "pushed" by TRG according to consumer subscriptions or "pulled" by protocols and applications with minimized retrieval delay in a distributed fashion. TRG glues together ANISI and ANHASA by registering unique universal context identifiers (UCIs) for multiaccess-related information with the ANISI ConCoord [2]. Then, all consumers can locate the multiaccess event sources using standard ANISI mechanisms. Storing multiaccess information in a way that addresses its dynamicity and heterogeneity is critical for ANHASA just as it is to make this information available to other entities. By distributing multiaccess information via TRG and ANISI, new multiaccess-aware applications and protocols can be introduced in a node without any modifications to lower-layer components (MRRM or GLL). This facilitates an open protocol stack and reinforces modularity as multiaccess information is disseminated without tying upper and lower protocols to any particular configuration.

*E. ANHASA Operation*

MRRM configures GLL to report periodically on new access availability. GLL uses link-specific scanning procedures, monitors the surroundings, and detects all available networks. Note that MRRM is RAT-agnostic and GLL is the only part of ANHASA that is technology-specific as it depends on the subjacent technologies; therefore, GLL is responsible for executing all requests coming from MRRM. MRRM can also issue a spontaneous request to GLL to detect new accesses upon reception, for example, of a trigger indicating that the currently used access network does not satisfy upper layer requirements. MRRM gathers information from the entire stack and decides which the most suitable access is. MRRM also initiates the processes of attachment to and detachment from a given access.

Once the multiaccess node has attached to a network, GLL supports MRRM operation by periodically providing abstracted link quality metrics from all active and available RATs. This enables MRRM to act swiftly and execute its access selection and handover algorithms when the connection deteriorates or a more suitable access becomes available.

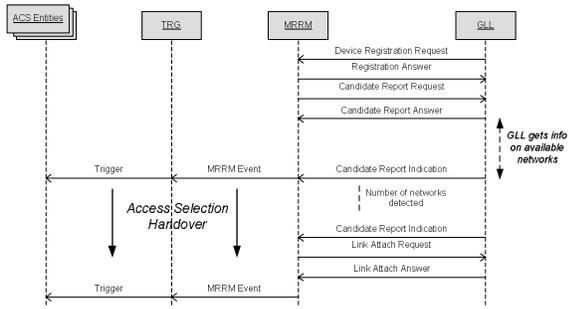

Fig. 3. ANHASA component interaction

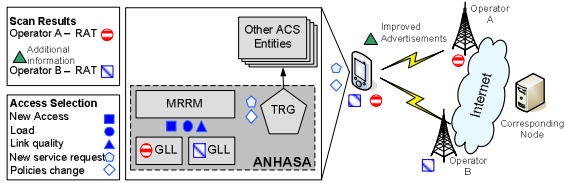

Fig. 4. Use case for ANHASA

Access details must be furnished promptly because of the time-varying characteristics of radio links. The frequency of updates is configurable and can be modified on demand. For instance, if we are running a real-time application, MRRM can instruct GLL to supply measurements every 100 ms. If a presence application or instant messaging is used instead, the update frequency can be smaller (say, every 500 ms). Clearly there is a tradeoff, which involves power consumption, system resources, and responsiveness, a topic of future study.

Fig. 3 illustrates how MRRM, GLL and TRG interact during access selection. First, MRRM and GLL cooperate to produce the "candidate (access) report", as explained above. This report may be of interest to entities within the node, and the associated event generates a trigger. This way, MRRM does not have to keep track of the upper layer entities interested in multiaccess details. On the other hand, TRG filters and correlates events according to consumer subscriptions and preferences, but does not take any decisive action related to mobility management or access selection (the domain of MRRM). Instead, TRG keeps the ACS up-to-date on any changes in the access networks, factoring out common functionality and allowing for incremental deployment of ANHASA, as we will discuss in §III.

*F. Access Selection*

Consider a user with a single terminal or a personal area network (PAN) supporting several RATs. The user is in an area covered by different operators, as illustrated in Fig. 4. The access selection algorithm considers different input parameters with pre-defined, but configurable weights as detailed in [10][11]. When the terminal is switched on, GLL will need to detect all available radio accesses. Since a full scan is costly in terms of time and battery power, GLL starts by scanning the RATs and frequencies used previously.

Access selection can also consider information stored in a SIM card or ANISI. Such pre-configured or slow-changing information, respectively, includes policies about the preferred operators and, possibly, which RATs to use under which conditions [12]. Ambient Network attachment [13] can also be used, when available: For example, Operator A in Fig. 4, may broadcast network capabilities and provide information about other available accesses in the area. Only if no accesses are found, a full scan is performed.

After establishing connectivity, further access selection(s) may take place by either MRRM-TE or MRRM-NET for terminal- or network-controlled multiaccess, respectively. Reasons for new access selection(s) include: detection of new access; coverage of one access is lost and a handover needs to be performed; new session initiations which require different QoS levels and other available RAT(s) may be more suitable than the currently selected. Additional hard termination criteria are also considered. For instance, if the cell load reaches a threshold, then the corresponding access(es) will not be further considered as a candidate.

Access selection is applied on a per-flow basis and can be broken into two stages: policy-based and dynamic. The former operates on a set of pre-defined parameters and (slow-changing) policies providing a ranked list of allowed candidate accesses. Policies embrace aspects such as security, roaming agreements, cost, and so on, and do not address the actual QoS requirements of any particular flow. When considering a network-controlled multiaccess environment, MRRM-NET needs information on the policies of the current accesses, and gathers it using TRG/ANISI. TRG is first informed about newly detected access networks or operators via a "Policies-Check-Request" MRRM event. In response, TRG returns a "Policies-Check-Answer". TRG can also trigger MRRM when security alerts occur, authentication certificates expire, changes in accounting due to prepaid account expiration, or due to a change in a composition agreement [14].

Dynamic access selection, on the other hand, operates on a set of time-varying parameters. MRRM determines whether the current access characteristics are sufficient to cope with the QoS requirements of all flows. MRRM-NET receives data about cell capabilities, capacity and load from the GLL on the network side. This includes the operational status of heterogeneous radio accesses, their current/average resource utilization and/or residual resources, enabling MRRM to remain up-to-date on network topology and operational status. Further, MRRM-NET receives measurements on the quality of the radio link that supports the flow and terminal based measurements of available candidate radio links. Thus, MRRM-NET obtains relevant information to perform dynamic access selection taking into account (a) the requested QoS by the user/application, (b) the radio link characteristics, (c) the network/cell capabilities and load, (d) the terminal capabilities, and (e) policy information.

To sum up, access selection generates a ranked list of all accesses and the highest ranked is used for each user flow. For an ongoing flow, if the chosen access is different from the one currently used, MRRM sends a handover execution re-quest as an event to TRG, which generates a trigger for mobility management protocols, tools, and other system components so that IP connectivity can be reestablished.

## III. ANHASA IN PRACTICE

This section discusses ANHASA realization. Simulation studies, such as [7], are critical in exploring large scale scenarios and validating the main concepts. Still, we argue that experimental work is required as well in order to assess the feasibility of the architecture in real devices and networks, and examine its actual overhead on a multiaccess node.

### A. Implementation and Testing

The main ANHASA components have been developed by several partners in the course of the two-phase/four-year Ambient Networks project as manifested by the references in this paper. Such a distributed development environment is well suited for the proposed architecture: agreements on service access points, interfaces, and signaling need to be reached, but then isolated component development and testing is possible. When components pass certain tests they are integrated into a single prototype. These two stages in the development of ANHASA are not sequential: insights from component testing feed into the specification while problems during integration expand the component testing and evaluation range. The end result is *running code* on real networks and multiaccess devices, not merely simulation models.

In terms of component testing, MRRM and GLL have been evaluated using Mobile IP-based mobility management. The testbed includes a dual-access (3G/UMTS and IEEE 802.11a) node running FTP, HTTP, and video streaming applications. The need for a handover is emulated with WLAN signal attenuation. When tuning the attenuator, MRRM-NET timely detects fading or improving candidate link conditions and proceeds with access selection, based on the (dynamic) abstract link quality reported by GLL and (static) preferences. MRRM was demonstrated [4][15] to successfully manage heterogeneous access selection while applications are running, without radio link disruptions or user intervention. Similarly, TRG was used to initiate lossless video streaming session handovers [8][16] and has proven an efficient means for gathering, filtering and delivering triggers requiring about one millisecond per trigger. More recently, TRG and GLL have been integrated in another demonstration [17], and we report the first lab measurement results in the following subsection.

It is important to recall that "MRRM", "GLL" and "TRG" refer to functional entities, not the actual implementations. The demonstrations indicate that ANHASA can be deployed incrementally, where nodes will incorporate individual components gradually and possibly in different configurations (MRRM+GLL, TRG-only, TRG+GLL, and so on). Nevertheless, the most benefits will be reaped only when all of ANHASA will be in place. For instance, the MRRM/GLL/MIP demonstration [15], while successful, does not let other parts of the protocol stack to become aware nor control the access selection process, and only static preferences are taken into account. Also, it is tightly integrated with Mobile IP and requires further tight integration if another mobility management protocol needs to be added in the stack. Similarly, the TRG demonstration, allows system components and orientation sensors to trigger access and device selection, but lacks genuine multiradio resource management.

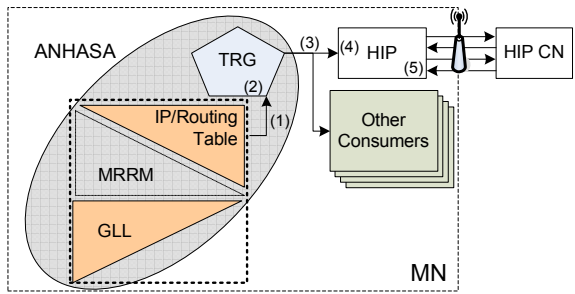

Fig. 5. ANHASA Lab Evaluation

### B. Lab Validation Results

We now present and discuss the first set of measurements from an ANHASA realization which includes TRG and GLL, but not MRRM. We consider a LAN to WLAN handover for the case of a HIP [18] Mobile Node (MN) and a Mobile Router (MR). Software and prototype details are provided in [17][19]. Initially, MN is streaming video from a HIP Corresponding Node (CN) over a LAN interface. Then, MN performs a HIP handover to the WLAN access point connected to MR. Finally, MR performs a handover from the LAN to another WLAN access point, emulating network mobility. Fig. 5 illustrates the relevant parts of the protocol stack for our tests. We insert trace points in our code to record the time spent in each phase of the handover process during testing. In addition the Wireshark network trace capturing tool is used to measure the handover and HIP signaling delay.

At the MN a handover is triggered based on an IPv6 Router Advertisement (RA) [20], which is captured at the WLAN-interface. Based on the RA, a new IPv6 address is autoconfigured on the WLAN interface. GLL sends this information to TRG (trace point (1) in Fig. 5). TRG processes the event (trace point (2)) and produces a trigger for HIP and other consumers (3). HIP processes the incoming trigger (4) and executes the handover after a four-way handshake with its CN (5). At the MR, a new IPv6 address and default router have been configured before initiating the LAN to WLAN handover, which is emulated by physically unplugging the cable. This (LAN) Link-Down event is delivered to HIP by a trigger originating from GLL.

The average processing delays at the software trace points are summarized in Table I. In both MN and MR cases, the total handover time is dominated by the four-way HIP update signaling, trace point (5). For the case of MR, this requires 300 ms, on average. In the case of a MN, significantly more time is needed (2.8s). For a MN handover, trigger delivery from GLL to HIP consumes 3 ms; in the MR case it requires 20 ms. A major difference between the MR and MN handover cases is the larger delay at trace point (1): 209 vs.

TABLE I ANHASA LAB TEST MEASUREMENTS

| Trace point | MN | MR |
|---|---|---|
| (1) | 209 ms | 10 ms |
| (2) | 2 | 1 |
| (3) | 1 | 19 |
| (4) | 13 | 16 |
| (5) | 2809 | 302 |

10 ms, on average, for MN and MR handovers, respectively. We need more detailed analysis to conclude why this is the case. At this stage, we attribute it to the handling of the RA at the MN: IPv6 auto-configuration and duplicate address detection introduces additional delay to the MN handover. These procedures have already concluded in the MR case. All in all, the MN handover requires approximately 3 s and the MR handover less than 350 ms, on average. Further trace points, more testing and analysis are in the works and will aid us in apportioning the delays to each component with greater accuracy, providing a detailed breakdown and opportunities for additional code optimizations.

## IV. RELATED WORK AND DISCUSSION

Standardization bodies, such as 3GPP [21], have been looking into multiaccess issues. For 3GPP, of central concern is the inclusion of different access technologies into reference scenarios. The work on Common Radio Resource Management (CRRM) [22][23] addresses multiaccess issues but focuses on handovers between the GSM/EDGE Radio Access Network (GERAN) and the UMTS Terrestrial Radio Access Network (UTRAN). CRRM is an example where inter-RAT or vertical handovers address two different 3GPP accesses. On the other hand, mechanisms for vertical handovers between wireless LANs and WANs have been studied for quite some time (see [24][25] and the references therein) but they typically rely on link availability events and other solely link-layer metrics to perform access selection and handover. ANHASA goes beyond "3G complemented by WLAN" scenarios paving the way for true multiaccess, where each flow can influence the access selection process.

The work in IEEE 802.21 [26], which defines media-independent mechanisms that optimize handovers between heterogeneous IEEE 802 and cellular systems is particularly pertinent to our effort. However, the three services it defines are not enough to cope with the relevant dynamicity and flexibility taken into account when specifying ANHASA. After all, access selection is out of the scope of the draft standard, which is simply an integration enabler for multiple RATs. So is multiradio resource management in subsecond timescales. Resource abstraction is a feature of 802.21, but not its implementation. Finally, the draft standard does not define a service that delivers upper layers events to the lower layers. Lower layers can be only "commanded" to take certain actions. Of course, once the standard is finalized, 802.21 compatibility could be a reasonable but additional feature for ANHASA.

Other collaborative projects, such as Daidalos, are also working on heterogeneous network architectures [27]. For instance, Daidalos defines an Abstraction Layer for different technologies, although they do not employ an event filtering and delivery service and, focus more on particular technological aspects. Although the targeted heterogeneous scenarios have gathered an increased interest from the research community recently, to the best of our understanding, there is no fully-fledged architecture similar to the ACS. ANHASA is an integral part of the ACS and thus benefits from all associated concepts including ANISI [2] and composition [14], simply to name a few.

## V. CONCLUSION AND FUTURE WORK

We presented ANHASA, an architecture that deals with the heterogeneity which appears as one of the most relevant characteristics of future wireless communications. It comprises three main functional entities, namely the generic link layer, multiradio resource management, and trigger management. ANHASA is designed with the intention to use the backend functionality provided by the Ambient Control Space when available. In this paper, we discussed, for example, how ANHASA can be integrated with the Ambient Networks information service infrastructure. ANHASA fully subscribes to the main ACS goal, which is to avoid the solutions patchwork currently dominating networks.

By emphasizing standardized APIs, ANHASA increases information sharing while respecting layering, the cornerstone of the TCP/IP stack. Each ANHASA component deals with a specific layer and does not meddle in a convoluted fashion with the rest of the protocol stack. For example, it allows MRRM on the lower end of the protocol stack to handle time critical events without hardwiring its use with any particular mobility management protocol. In ANHASA, MRRM can easily be employed in other systems that favor, say, HIP instead of Mobile IP. ANHASA is inclusive, as it supports different implementations of the constituent functional entities. It is generic, and not only an optimization aiming for particular scenarios, such as other cross-layer solutions for network multimedia.

ANHASA components have already been integrated, as parts of the ACS, into a single prototype. ANHASA components have also been used in public demonstrations, showcasing seamless vertical handovers and video streaming session handovers, with different mobility management protocols. The results reported in this paper are very promising, indicating that ANHASA is a viable solution. Yet, however successful these demonstrations and lab tests may be, they are just checkpoints towards realizing the complete ANHASA/ACS.

Our near-future plans include the final integration of all ANHASA components in a single prototype. With the inclusion of MRRM to the TRG and GLL prototype presented in §III.B we will be able to realize, for example, mobility management with different protocols using dynamic access selection. The experimental framework we presented will be expanded as well, to include more ACS functional entities. More detailed analysis of the delays and overheads required to perform access selection under various circumstances is in our plans too. Furthermore, we will employ simulations to explore the benefits of ANHASA in large-scale scenarios.

Last but not least, the integration of ANHASA with ANISI and policy management will need to proceed further.

ACKNOWLEDGMENT

This work has been carried out in the framework of the Ambient Networks project (IST 027662), which is partially funded by the Commission of the European Union. The views expressed in this paper are solely those of the authors and do not necessarily represent the views of their employers, the Ambient Networks project, or the Commission of the European Union. The comments and ideas from people involved in the project's multiaccess and mobility research are gratefully acknowledged.